\documentclass{article}

{}
{}
{}

\usepackage{multirow}
\usepackage{color}
\newcommand{\review}[1]{{\textcolor{black}{#1}}}

\usepackage{amsmath} 
\usepackage{amsthm} 
\usepackage{cite} 
\usepackage{hyperref} 
\usepackage{graphicx}
\usepackage{algorithmic} 
\usepackage{hyperref}
\usepackage{authblk}
\usepackage[margin=1.2cm]{geometry}
\usepackage{multicol}
\usepackage{blindtext}
\usepackage{epstopdf}
\begin{document}



\title{\textbf{Exploring the Risks and Challenges of National Electronic Identity (NeID) System}}

\author[1]{Jide Edu}
\author[1]{Mark Hooper}
\author[1]{Carsten Maple}
\author[1]{Jon Crowcroft}

\affil[1]{{The Alan Turing Institute}}

\date{}

\setcounter{Maxaffil}{0}
\renewcommand\Affilfont{\itshape\small}

\maketitle

\begin{abstract}
Many countries have embraced national electronic identification (NeID) systems, recognising their potential to foster a fair, transparent, and well-governed society by ensuring the secure verification of citizens' identities. 
The inclusive nature of NeID empowers people to exercise their rights while holding them accountable for fulfilling their obligations.
Nevertheless, the development and implementation of these complex identity-verification systems have raised concerns regarding security, privacy, and exclusion. 
In this study, we discuss the different categories of NeID risk and explore the successful deployment of these systems, while examining how the specific risks and other challenges posed by this technology are addressed.
Based on the review of the different NeID systems and the efforts made to mitigate the unique risks and challenges presented within each deployment, we highlighted the best practices for mitigating risk, including implementing strong security measures, conducting regular risk assessments, and involving stakeholders in the design and implementation of the system.

\end{abstract}

\begin{multicols}{2}

\section{Introduction}

Governments worldwide are committed to supporting the rollout of the national electronic identity (NeID) system to improve efficiency and ensure the proper delivery of welfare and benefits to citizens. These identification systems are crucial for sustainable development, and have helped empower citizens by enhancing their access to rights, services, and the formal economy. Governments use these systems to streamline public administration, improve security, and provide improved services \cite{Joseph_2018}. 
Financial institutions also use NeID systems to verify their customers' identity, which addresses a vital issue of Know Your Customer (KYC) and enables them to provide a wide range of financial services, including opening accounts, securing credit, taking deposits, and paying for services. 
The exercise of fundamental human rights, claim of entitlements, access to a range of government services, and conduct of many daily activities can be hindered without reliable identification.

However, the development and implementation of NeIDs also raise concerns about security, privacy, and exclusion. For example, individual data are sometimes concentrated in a centralised system, and there is potential for data breaches and misuse.
Several incidents have already been reported involving NeID systems, such as the Aadhaar case, in which over 200 government websites exposed data of Indian citizens \cite{Tiwari_and_Agarwal_2022}.
%
These systems are attractive to adversaries because they contain considerable amounts of personal data \cite{Aadhaar_system}. 
For instance, the World Bank Group's ID4D initiative identified thirty-five possible threats to identity system security and privacy \cite{ID4D_threats}.

More importantly, the NeID system depends heavily on technology and comes with all risks associated with complex technical programs \cite{Joseph_2018}.
People are becoming reluctant to provide personal data because of the fear of identity theft or data misuse \cite{10.1007/978-3-642-03547-0_18, 10.6028/NIST.SP.800-63-3} and many are worried about privacy and government surveillance concerns \cite{Halperin_2012}.
As governments and organisations begin to deploy and use NeID systems to provide digital services and improve efficiency, they are exposed to the risks associated with these systems. In this study, we examine the associated risks and challenges of the NeID system and how they are being addressed.
We begin by examining the history of NeID development, NeID key components, and the implementation approaches adopted by various countries (Section \ref{sec:Back}). We discuss in Section \ref{sec:riskandchallenges} the risks and challenges associated with the NeID System before diving into the best practices for mitigating the risks, examples of successful deployments, and how these risks have been addressed (Section \ref{sec:mitigating}).



\section{Background and Context}
\label{sec:Back}

\subsection{History of NeID systems}

NeID systems have evolved significantly over time, intending to create more secure, easy-to-use, and efficient ways of verifying a person's identity. 
The first NeID system, Estonian eID, was introduced in Estonia in 2002 \cite{martens2010electronic}. It allows citizens to digitally sign documents, authenticate their identities online, and access government services. This system has been a significant success in Estonia, with over 98\% of the population regularly using it \cite{martens2010electronic}. Nevertheless, there remains a need for widespread adoption, user education and awareness, and secure infrastructure \cite{Moon_2017}.

Singapore was another early adopter of NeID systems. In 2003, the Singaporean government launched the National Authentication Framework (NAF), allowing citizens to authenticate their identities online using a single login credential \cite{SinglePass}. 
This provides Singapore citizens and residents with a convenient way to access government services online, including tax filings, healthcare appointments, and banking transactions.
In 2018, the government launched SingPass Mobile \cite{SinglePass}, a mobile application that enables citizens to access government services using their smartphones.
%
%
Equally, in 2011, India launched its NeID system, called Aadhaar \cite{Aadhaar_system}. Aadhaar provides Indian citizens with a unique 12-digit identification number that can be used to access various government services, including social welfare benefits and healthcare \cite{Tiwari_and_Agarwal_2022}. 
Nonetheless, the system has also been criticised for its privacy issues and allegations of misuse \cite{Aadhaar_2018}.

The United Kingdom launched its NeID system, GOV.UK Verify \cite{GOV.UK} in 2016. GOV.UK Verify allows citizens to verify their identity online when accessing government services. However, the system has been criticised for being difficult to use and for its low adoption rate.  \cite{Gov_Verify}.
More recently, other countries have begun to develop NeID systems. For example, Australia launched a digital identity system called myGovID in 2019 \cite{hanson2018preventing}. This system allows citizens to access government services online through smartphones or computers. However, while the system has successfully improved accessibility to government services, it has also been criticised for being too reliant on mobile technology and not offering enough support to people without smartphones or reliable internet access \cite{hanson2018preventing}.


\subsection{NeID components}
NeID Systems work by capturing each individual's unique identity, which is collated in a centralised or decentralised identity database, with each individual having a unique identifying number (UIN) \cite{Joseph_2018}. A government may issue official identification credentials such as national identity cards, and operate identity services that verify a person's identity online. 
%
The digital ID process consists of three main components: i) identity proofing and enrolment, ii) authentication and identity lifecycle management, and iii) portability and interoperability mechanisms \cite{FATF_2020}. These components highlight the complex and dynamic nature of the NeID systems. 

\subsubsection{Identity proofing and enrolment}
The first phase of NeID involves identity verification and enrolment. This answers the question of who an individual is by collecting, validating, and verifying information about them within a given population or context. These may be digital, documentary, in-person, or remote \cite{FATF_2020}. 
The main objective is to identify and verify an individual and bind their identity to an authenticator. This phase is further divided into four:


\noindent\textit{\textbf{Collection and resolution:}} 
In this phase, attributes are obtained, evidence of attributes is gathered, and identity evidence and attributes are resolved into a distinct identity within a particular population or context.
There is a de-duplication process that involves checking the individual applicant's biographic attributes, biometrics, and attributes assigned by the government, such as driver's license, or passport numbers, against the database of enrolled individuals, as well as the attributes and identity evidence associated with them, to avoid duplicate enrolment.

\noindent\textit{\textbf{Validation:}} 
This phase involved verifying the accuracy and authenticity of the collected evidence. For instance, the identity provider (IDP) can examine the applicant's physical identities evidence, such as a driver's license or passport, and determine that they have not been altered, that the digital security features are intact, and that the information on the physical identity evidence matches the sources.

\noindent\textit{\textbf{Verification:}} This entails confirming that the applicant's identity is associated with a validated identity. As part of the liveness checks, the IDP may, for example, request that the applicant submit a photo or video from their mobile device. Photographs of passports, licenses on government databases can be compared to an applicant's submitted photo with some degree of certainty \cite{Joseph_2018}. A valid phone number linked to the applicant's identity can then be sent an enrolment code that the applicant must provide to the IDP. This code must match that sent by the IDP, proving that an applicant is a real person who owns and controls the validated phone number \cite{FATF_2020}.

\noindent\textit{\textbf{Enrolment in identity account and binding:}} At this stage, the identity account is created, and one or more authenticators are linked to the identity account, allowing for identity authentication.

\subsubsection{Authentication and ID lifecycle management}
Authentication determines whether a person is who they say they are.
This establishes that the person asserting identity is the same person who was verified and enrolled. Three types of factors can be used to authenticate an individual \cite{FATF_2020}: i) factors of ownership (something you own, such as cryptographic keys), ii) factors of knowledge (something you know, like a password), and iii) inherent factors, such as biometrics or something you are.
In addition, identity lifecycle management refers to the actions IDPs should take in response to events that can occur during the lifecycle of a subscriber's authenticator. These events affect the use, security, and trustworthiness of the authenticators. Some examples of these could be issuing and binding authenticators to credentials, either during or after enrolment. IDPs can also take actions relating to expiration, revocation, loss, theft, and unauthorised duplication of authenticators or credentials.

\vspace{-2mm}
\subsubsection{Portability and interoperability mechanisms}
Through portability, people can identify themselves across multiple networks using their digital ID credentials, rather than repeating the process of obtaining and verifying personal information.
One approach to enabling a portable ID is through federation. The federated architecture and assertion protocols allow the transmission of identity and authentication data between the networked systems, enabling different networks to work together. For example, GOV.UK \cite{GOV.UK} provides an excellent example of a federated digital ID, where functions and responsibilities are spread across service providers.

\vspace{-2mm}
\subsection{NeID systems implementation}
The implementation of NeID systems varies by country. However, they can be classified into two main categories: centralised and decentralised systems \cite{dib2020decentralized}.

\noindent\textbf{\textit{Centralised}}: Participants in centralised system rely on the platform operator's trustworthiness to conduct transactions successfully, securely, and privately. Therefore, the trust relationship is mandatory.
The Aadhaar system in India is an example of a centralised digital ID system. 
Over 1.2 billion people's biometric and demographic information is stored in the Aadhaar centralised database.
Another example is Nigeria, which has implemented a centralised NeID system called the National Identification Number (NIN) \cite{ayamba2016national}.
All Nigerian citizens and legal residents are required to have NIN. 
A Nigerian National Identity Card is issued upon completion of enrolment into the Nigerian National Identity Database (NIDB) and 
consists of 11 digit numbers. The database is tied to essential public services to improve management and efficiency. Countries such as Ethiopia, Singapore, Malaysia, and South Korea also use centralised digital ID systems. One of the key potential benefits of a centralised digital ID is its apparent authority to appeal to. This occurs in situations where the use of a digital ID is disputed. Other benefits include increased efficiency in service delivery and enhanced identity verification for government and private sector activities.
However, there are also concerns about privacy and the potential misuse of personal data in centralised systems \cite{Dunstone_2009, 10.1007/978-3-642-03547-0_18, 10.6028/NIST.SP.800-63-3}.

\noindent\textbf{\textit{Decentralised}}: In a decentralised system, each user is responsible for their own identity \cite{dib2020decentralized}. A good example is Estonia's eID system. This system is based on a decentralised architecture that allows citizens to securely access government and private sector services using a single identity credential known as eID \cite{martens2010electronic}.
eID is a smart card containing a microchip with a unique digital identity certificate linked to the citizen's identification code. eID allows citizens to access various online services securely, including voting, banking, healthcare and taxation.
Another country with a decentralised NeID is Switzerland, which launched a system known as SwissID \cite{mettler2019suisseid}. SwissID is a joint initiative among several Swiss companies and institutions. It is based on a blockchain-based architecture that allows users to control their data and manage their identity credentials securely.
The apparent strengths of decentralised systems include increased privacy, security, user control, and improved interoperability across different sectors. However, decentralised NeID systems introduce additional technological complexities and are often challenging to scale compared to the centralised approach \cite{chu2018curses}.


\section{Risks and Challenges}
\label{sec:riskandchallenges}
Although various deployments of NeID systems exist across countries, they perform similar functions and share common features and risks. In particular, NeID systems support individual registration, personal data collection, credentials management, and online authentication and authorisation. Next, the risks associated with these systems are discussed.

\subsection{Security and privacy concerns}

\noindent\textbf{\textit{Impersonation risk}}: 
The risks of presenting false evidence (either stolen or counterfeit) can be actualised on a much larger scale in NeID systems \cite{Dunstone_2009}. Impersonation involves a person pretending to have the identity of genuine person. Although this type of attack does not constitute a large-scale hack if it affects only a single individual or a few individuals, it may nonetheless significantly impact the victims \cite{Remote_ID_Proofing}.

\noindent\textbf{\textit{Synthetic IDs}}:
Criminals develop synthetic identities by combining real (usually stolen) and fake information to create a fictitious identity \cite{Dunstone_2009} that can be used to open fraudulent accounts and make purchases. Unlike impersonation, the criminal is pretending to be someone who does not exist in the real world rather than impersonating an existing identity. Synthetic IDs can be used to obtain credit cards or online loans and withdraw funds with accounts abandoned shortly thereafter. An excellent example of synthetic ID fraud is the 2018 Aadhaar system hacks \cite{Khaira_2018}, in which software vulnerability allows the generation of unauthorised Aadhaar numbers.

\noindent\textbf{\textit{Data corruption, error, and loss}}:
The reliability of NeID can be undermined by source documents that are simple to tamper with. For example, massive attack fraud is more likely to occur remotely than in person or through processes that require human intervention \cite{FATF_2020, Remote_ID_Proofing}. Although NeID alleviates some drawbacks of paper-based systems by providing security features and secure authentication, it also poses a major risk of data corruption, error, and loss. For example, data may be introduced, modified, or deleted without authorisation.

\noindent\textbf{\textit{Identity life cycle risks}}: 
An inadequate identity lifecycle and access management can intentionally or unintentionally compromise the authenticity of authenticators and enable unauthorised individuals to access and misuse the privileges of ID users \cite{Remote_ID_Proofing}. For example, an identity must be revoked or terminated promptly when it ceases to exist (such as when the identity owner dies), is found fraudulent, or no longer meets the eligibility requirements. Failure to implement better identity lifecycle management exposes the system to risk, as a malicious party can exploit such an ID.


\noindent\textbf{\textit{Authentication risks}}:
Authentication risks occur when a legitimately issued digital ID is compromised, and its credentials or authenticators are under the control of an unauthorised person. Malicious actors may assert an individual's legitimate identity to a relying party to obtain unauthorised access to services and data because of vulnerabilities in the types and number of authentication factors \cite{Dunstone_2009, Remote_ID_Proofing}. 
For instance, biometric data can be stolen in bulk from central databases. An example is the Philippines election hack \cite{Leisha_2016}, in which the data of over 70 million registered voters were compromised.
%
Unlike passwords, which can easily be changed if compromised, individual possesses only a finite number of biometric attributes (2 eyes, 10 fingers, and a face), which are difficult to replace.

\noindent\textbf{\textit{Data breach}}:
NeID systems hold massive PII, including biometrics belonging to a considerable number of people, making them an interesting target for adversaries.
In the context of identity theft, it is impossible to guarantee the security of such vast databases \cite{LSE_ID_report}. 
A successful attack on such a database can compromise the massive PII of individuals \cite{BBC_2014} and potentially lead to large-scale identity theft \cite{Leisha_2016}. 
These could pose major privacy breaches, fraud, or other related financial crime risks, owing to the potential scale of the attack.

\noindent\textbf{\textit{Open communications network}}:
Identity-proofing and authenticating individuals over an open communication network (the internet) creates additional risks for NeID systems.
It presents multiple cyberattack opportunities between the communicating parties (ID Provider, relying party, and end-user) \cite{FATF_2020}. Criminals and other bad actors may be able to abuse the systems to create false identities or exploit authenticators linked to a legitimate identity.

\vspace{-2mm}    
\subsection{Digital divide and exclusion}
The NeID System can exacerbate existing inequalities and create a digital divide between those with access to the system and those without access. For example, when NeID systems do not exist, cover all or most persons in a jurisdiction, or exclude specific populations, they may drive the exclusion of services \cite{FATF_2020}. 
Biometric failures, Internet connectivity issues, lack of accessibility or reliable infrastructure can undermine the NeID system 
excluding residents from accessing critical services.
In India, there have been reports of starvation deaths caused by the failure to access critical government services \cite{Dey_2018}.
There is also the risk of exclusion when countering fraud \cite{Dreze_2017}. Moreover, bias in biometric technology can disproportionately affect certain races and genders \cite{Buolamwini2018GenderSI}. Exclusion can also occur when an ID system is too complicated for users to understand or relies too heavily on obscure technology \cite{4489846}.

\subsection{Legal and regulatory issues}

\noindent\textbf{\textit{Data protection risk}}: NeID systems collect and store a large amount of personal data, including biometric data, which raises concerns about privacy and data protection. The lack of adequate privacy safeguards could expose citizens' data to risk.
Additionally, NeID may affect people's ability to exercise their privacy. For example, in India, women who had previously saved secretly were forced to open bank accounts, putting their savings at a greater risk of exposure to spouses and family members who had power over them \cite{Beduschi_2017}.

\noindent\textbf{\textit{Interoperability}}: NeID systems need to be interoperable with government and private sector systems to deliver seamless service. In this way, non-compatibility with the relying organisation’s policies, industry regulations, legal framework, and standards could hinder interoperability and adoption.

\noindent\textbf{\textit{Data misuse}}:
A major risk associated with digital identity adoption is how the government and private sector will use and store biometric information \cite{Halperin_2012}. Profiling, surveillance, intelligence gathering, and data sharing without user consent are other potential misuses associated with centralised foundational ID systems \cite{Dunstone_2009, 10.1007/978-3-642-03547-0_18, 10.6028/NIST.SP.800-63-3}. This is most likely to occur if there are not enough privacy and data protection laws in place, if those laws are not enforced well, or if a new law is passed that provides access to the database for other purposes after the system has been implemented. 
%
Similarly, there is a possibility of abuse when NeID systems are made available for commercial applications.
A case study is India \cite{Aadhaar_2018} where it was ruled that Aadhaar has been used illegally for verification during the creation of bank accounts, and mobile numbers.

\section{Discussion}
\label{sec:mitigating}

\subsection{Case studies of successful deployments}

There have been several successful implementations of NeID systems worldwide, each with unique features and approaches to address risks and challenges. Next, we discuss some examples of such implementations. 

{\noindent\textbf{\textit{India's Aadhaar}}: India's Aadhaar system has been hailed as one of the largest NeID implementations in the world, with over 1.2 billion registered users.
%
It uses multifactor authentication, which combines biometric authentication (fingerprints and iris scans), a one-time password (OTP), and demographic authentication to ensure that only an authorised person can access their Aadhaar information \cite{Tiwari_and_Agarwal_2022}. Aadhaar also uses a QR code containing encrypted information that authorised parties can only read to provide additional security for user data.
It uses virtual IDs (VIDs), which are temporary numbers that can be used instead of Aadhaar numbers for authentication purposes \cite{Tiwari_and_Agarwal_2022}. This helps protect the Aadhaar number from being shared with unauthorised people. Aadhaar data are also restricted to only authorised organisations and use advanced encryption to protect the collected personal data \cite{Tiwari_and_Agarwal_2022}}.


{Aadhaar system is available in multiple languages and is designed to be user-friendly, with a simple interface to ensure that all users can access it easily \cite{Tiwari_and_Agarwal_2022}. Furthermore, Aadhaar provides alternative verification methods for users who may not have access to a mobile phone or the Internet, such as an Aadhaar letter, that can be used for authentication. 
%
Nevertheless, there is still a need for clear and comprehensive data protection laws to protect individual privacy and personal data. While the Personal Data Protection Bill was introduced in 2019 to address this issue, it is yet to be enacted \cite{PDPB_2022}. 
There have also been several legal challenges to the Aadhaar system, with critics arguing that it violates citizens' right to privacy and can be used for surveillance purposes \cite{Aadhaar_2018}. 
Aadhaar also suffers from limited accessibility and the risk of exclusion, especially in rural areas with limited technology access.} 

{\noindent\textbf{\textit{Singapore's SingPass}}: Singapore's NeID system, SingPass, uses multifactor authentication to verify the identity of its users \cite{SinglePass}. The system requires users to provide a username and password, biometric verification, and a One-Time Password (OTP) sent to their registered phone number or generated by a security token \cite{SinglePass}. 
All communication between the user's device and the SingPass server is encrypted using the Transport Layer Security (TLS) protocol, which provides high security for online transactions \cite{cooper2022national}. Moreover, users are given control over their data. For example, users can view their login history, change their passwords, personal information, and set up notifications for suspicious activities such as login attempts from unknown devices or locations.}

{SingPass is designed to be user friendly and accessible in multiple languages.
It is accessible to all users, including those with disabilities, and offers options for users with visual impairments such as screen readers and text-to-speech capabilities \cite{cooper2022national}. There is an alternative verification method for users who may not have access to a mobile phone or security token, such as a physical token that generates a one-time password.
%
The SingPass eID system operates based on the principle of consent, which means that users must give explicit consent before their data are collected, processed, or shared \cite{cooper2022national}. In addition, the SingPass user agreement explicitly states the purpose for which the data will be used and allows users to opt out.
Singapore has enacted various laws to support the SingPass eID system, including the Electronic Transactions Act (ETA), 
the Personal Data Protection Act (PDPA),
and the Cybersecurity Act, which establishes a framework for protecting critical information infrastructure.
SingPass is interoperable with other eID systems worldwide, allowing users to access online services in different countries and enabling users of other eID systems to access Singaporean online services.}

{However, the SingPass authentication method may exclude less-tech-savvy people who have difficulty remembering their passwords or navigating the authentication process. Likewise, SingPass is a single point of failure for accessing multiple government e-services. Therefore, any issue with the system could disrupt access to all the e-services that rely on it.}

{\noindent\textbf{\textit{Estonia's eID}}: Estonia is considered to have one of the most advanced NeID systems in the world \cite{cooper2022national}.
Estonia's eID system uses a two-factor authentication mechanism that requires a smart card and a PIN code. 
Users has complete control over the card 
and can decide which data to share with whom and revoke access when desired \cite{oecd2019digital}. In addition, the system uses advanced encryption techniques to secure communication between the user's device and online service provider. This ensures that sensitive information is protected from interception and unauthorised access. 
Furthermore, the Estonian government implemented a centralised security management system to monitor and detect any security breach, allowing a quick response to security issues.}

{Legislation has been enacted to support the system, including the Identity Documents Act, Personal Data Protection Act, Electronic Identification and Trust Services for Electronic Transactions Act, and the Digital Signatures Act.\footnote{https://www.skidsolutions.eu/en/repository/legislation} These laws provide a legal framework for the eID system and ensure it operates per European Union (EU) data protection and privacy regulations. 
In addition, Estonia's eID system is designed to be interoperable with other eID systems across the EU. This allows Estonian citizens to access online services in other EU countries and also allows citizens of other EU countries to access Estonian online services \cite{cooper2022national}. 
This system uses public-key cryptography, which provides a high level of security and ensures the integrity of electronic transactions.}

{Estonia's eID system is accessible to all citizens and residents regardless of their location or socioeconomic status. 
It is also designed to be user friendly and available in multiple languages, including Estonian, Russian, and English, to ensure that all users can easily access it. 
%
However, similar to SingPass, Estonia's eID system is a single point of failure for accessing multiple e-services. Also, there is a risk of limited accessibility, as the current deployment may be challenging for less-tech-savvy users with difficulty using an ID card reader.} 


\subsection{ Best practices for mitigating the risks} 

\noindent\textbf{\textit{Risk assessment}}:
For a NeID system to become a critical strategic asset, it must be widely used, highly trusted, reliable, and accessible \cite{Joseph_2018}. To achieve this, identifying and assessing NeID systems risks are crucial. This can be achieved through a risk assessment process that involves identifying potential threats, evaluating their likelihood, assessing their impacts, and prioritising them based on their level of risk. 
However, the risk assessment needs to consider the varied risk impacts on the different NeID stakeholders, the context of use and the community the NeID systems serve.
By understanding these risks, we can better prepare to take action to mitigate them.

\noindent\textbf{\textit{Stakeholders involvement}}:
There is also a need to involve the stakeholders in the design and implementation of the system. This will bring diverse perspectives and insights, ensuring that broader stakeholder concerns are considered as the system is being developed. 
Stakeholder involvement can help identify high-risk areas that require more attention and enable the development of more effective risk management strategies. It could also increase transparency and accountability, and build trust between stakeholders and the NeID provider.

\noindent\textbf{\textit{Privacy protection}}:
\review{Privacy generally refers to norms and practices that safeguard human autonomy, identity, and dignity. The freedom of individuals to consent to the disclosure or control aspects of their identities (such as their body, data, and reputation) and the limitations of observation are typically addressed by these norms and practices \cite{nissenbaum2009privacy}. 
Personal data are one of the most valuable assets of NeID. Therefore, privacy values like anonymity, confidentiality, and control should be prioritised for NeID system design, development, and deployment. 
Robust legal and regulatory frameworks will also ensure that NeID systems are effectively and transparently governed, as they will provide clear rules and regulations to manage personal data collection, use, and sharing.}



\section{Conclusion}
\label{sec:conclusions}

The evolution of NeID systems has been driven by the need for more secure and convenient ways to authenticate identities online. These systems support individual registration, personal data collection, credentials management, and online authentication. Despite its potential benefits, NeID has several technical challenges and risks that must be addressed. In this study, we discussed the different categories of NeID risk, and explored the successful deployment of these systems, whilst examining how specific risks and other challenges posed by this technology are addressed. In reviewing of different NeID systems and the efforts made in order to mitigate unique risks and challenges presented within each deployment, our findings show that successful implementations share common features, such as strong authentication measures, robust data protection and encryption protocols, and clear policies and regulations. As NeID systems become more widespread, it is imperative that all NeID stakeholders have a comprehensive understanding of the associated risks and challenges for the effective design and implementation of robust control measures to help make the system secure, accessible, reliable and trustworthy, thereby realising more benefits to citizens.

\vspace{-1mm}
\section{Acknowledgements}
This work was supported in whole 
by the Bill \& Melinda Gates Foundation [INV-001309]. Under the grant conditions of the Foundation, a Creative Commons Attribution 4.0 Generic License has already been assigned to the Author's Accepted Manuscript version that might arise from any submission.

\vspace{-3mm}

\end{multicols}

\end{document}